\documentclass[a4paper, twocolumn]{article}
\usepackage{graphicx}
\usepackage{courier}
\usepackage{flushend,cuted}
\usepackage{amsmath}
\usepackage{url}

\title{Exfiltration of Data from Air-gapped Networks via Unmodulated LED Status Indicators}
\author{Zheng Zhou, Weiming Zhang, Zichong Yang, Nenghai Yu\\
zhou7905@mail.ustc.edu.cn; zhangwm@ustc.edu.cn;\\  zcyang91@mail.ustc.edu.cn; ynh@ustc.edu.cn\\
University of Science and Technology of China\\
Key Laboratory of Electromagnetic Space Information\\ 
of Chinese Academy of Sciences} 
\begin{document}
\maketitle
\begin{strip}
\begin{abstract}
The light-emitting diode(LED) is widely used as an indicator on the information device. Early in 2002, Loughry et al studied the exfiltration of LED indicators\cite{Loughry:2002:ILO:545186.545189} and found the kind of LEDs unmodulated to indicate some state of the device can hardly be utilized to establish covert channels.
In our paper, a novel approach is proposed to modulate this kind of LEDs. 
We use binary frequency shift keying(B-FSK) to replace on-off keying(OOK) in modulation.
In order to verify the validity, we implement a prototype of an exfiltration malware.
Our experiment show a great improvement in the imperceptibility of covert communication.
It is available to leak data covertly from air-gapped networks via unmodulated LED status indicators. 
\end{abstract}

\begin{flushleft}
\textbf{Keywords: } Air-gapped Networks;\quad Covert Channel;\quad Light-Emitting Diode;\quad LED indicator;\quad Modulation
\end{flushleft}
\end{strip}
\raggedend
\section{Introduction}
The covert channel is a well-known way to transmit messages by circumventing the security mechanism. The definition of covert channel was given by Lampson in 1973 to describe the leakage of data by abuse of shared resource by the processes in different privilege levels\cite{Lampson:1973:NCP:362375.362389}. 
With the development of communication technology, the border of covert channel had been extended from one-host to networks. There are many kinds of cover channel developed in past twenty years. Zander et al surveyed the network covert channels in different kinds of networks protocols\cite{Zander2007-4317620}. In order to maintain the security, physical isolation is applied in almost every top-secret organization to keep the networks with high level separated from the less secure and public networks. The term of this type of isolation is \textit{air-gapped}.
Is the air-gapped networks safe enough then? No. A lot of methods were proposed to breach the air-gapped networks in the last ten years. Generally saying, there are four kinds of covert channel to bridge the air gap: \textit{Electromagnetic} covert channels, \textit{Acoustic} covert channels, \textit{Thermal} covert channels and \textit{Optical} covert channels.

Kuhn and Anderson proposed firstly the method\cite{kuhn1998soft} to transmit information covertly using electromagnetic radiation in 1998. 
Guri et al introduced AirHopper\cite{guri2014airhopper}, a type of malware, leak data between a mobile phone and a computer nearby using FM radio module in 2014. 
Guri et al introduced a malware named GSMem\cite{guri2015gsmem}, which leak data via electromagnetic radiation generated by the bus of computer memory in 2015.
Guri et al proposed USBee\cite{guri2016usbee}, which can be used to leak data via electromagnetic radiation generated by the USB cable in 2016.
In 2016, Matyunin et al used the magnetic field sensor in mobile device to build a covert channel.

In 2013, Hanspach and Goetz used the acoustical devices: speakers and microphones of the notebook computer to build a covert channel\cite{hanspach2014covert}.
Malley et al\cite{Malley-o2014bridging} introduced a covert communication over inaudible sounds in 2014.
Lee et al\cite{lee2015various} uses a loud-speaker as an acoustical input device, and make a speaker-to-speaker covert channel in 2015. 
Guri et al introduced Fansmitter\cite{Guri-Fansmitter-2016arXiv160605915G} and DiskFiltration\cite{Guri2017DiskFiltration}, new methods to send acoustic signals without speakers in 2016.

In 2015, Guri et al introduced BitWhisper\cite{guri2015bitwhisper}, to build a unique bidirectional thermal covert channel via the heat radiated with another adjacent PC. 
In 2017, Mirsky et al proposed HVACKer\cite{Mirsky2017}, to build a one-way thermal covert channel from an air conditioning system to an air-gapped network.
The thermal covert channels in the multi-cores CPU is researched as follows.
Mast built a thermal covert channel in multi-cores\cite{masti2015thermal} with a transmit rate of 12.5bits per second in 2015.
Bartolini studied the capacity of a thermal covert channel in multi-cores\cite{bartolini2016capacity} in 2016.
Selber propose UnCovert3\cite{selber2017uncovert3}, a new thermal covert channel in multi-cores with a transmit rate of 20 bits per second in 2017. 

The optical covert channels are mostly utilized. 
Shamir present a cover channel to breach an air-gapped network \cite{shamir2014light} by a light-based printer in 2014.
Lopes and Aranha proposed a malicious device\cite{lopes2017platform} to leak data via its flickering infrared LEDs.
In 2016, Guri introduced VisiSploit\cite{}, a prototype to leak date via an invisible QR-code in LCD screen.

Loughry and Umphres studied the exfiltration via LED indicators\cite{Loughry:2002:ILO:545186.545189} in 2002. They divided LED indicators into three classes:
\begin{description}
\item[Class I] The unmodulated LEDs used to indicate some state of the device.
\item[Class II] The time-modulated LEDs correlated with the activity level of the device.
\item[Class III] The modulated LEDs that are strongly correlated with the content of data being processed.
\end{description}
They found that TD LED indicators on almost every modem of those years belong to Class III. Even a LED indicator on a DES encryptor leaks plain data. They indicated that although the LEDs in Class II are not so dangerous as those in Class III, but they can be modulated to leak significant signal, and can be used to build covert channels.

Sepetnitsky proposed a covert channel prototype \cite{Sepetnitsky-2014-6975588} of leaking data to the camera in a smart phone via the monitor's power status LED indicator in 2014.
Guri presented LED-it-GO\cite{Guri2017LED}, to leak data via hard drive LED indicator in 2017.
Guri also proposed xLED\cite{guri2017xled}, to leak data via status LED indicators on the routers in 2017.

In Guri's two methods, LED-it-GO and xLED, the LEDs that used as the light source belong to Class II. They flickers naturally without causing user's suspicion. Nevertheless, Sepetnitsky's prototype might be not so good to cope with the behavior covertness of covert channel. Because the LED indicator he used belongs to Class I. Unfortunately, the fastest flicker frequency of the monitor power LED is 25Hz. It is hard to circumvent human sense of sight if some data is modulated with OOK at that frequency.

In our paper, a novel approach is proposed to modulate the LEDs in Class I. We give a prototype, KLONC(the abbreviation of ``Keyboard's LED tO Network Camera"), to build an optical covert channel and to leak data from air-gapped network to an IP camera via the LED status indicator on the keyboard of a PC. In 2002, Loughry et al presented an exfiltration via keyboard LED indicators in Appendix A\cite{Loughry:2002:ILO:545186.545189}. The flicker frequency was up to 150Hz in Solaris OS. Unlikely, because of the limitation of Windows 10, the ordinary user-leveled program can make the keyboard LED indicators flicker at 33Hz only by simulation of striking the keyboard. In our experiment, we noted that human vision can hardly distinguish two flickers with different frequencies on a LED lighting on continuously. Then, we use B-FSK to modulate the data. Two different flicker frequencies are utilized to encode logical '1' and  '0'. 
The result of experiment shows that the effect of covertness is achieved.
Our approach can be used in the optical covert channels via LED indicators in Class I at a low flicker frequency. Especially, Sepetnitsky's prototype\cite{Sepetnitsky-2014-6975588} can use our approach by replace OOK into B-FSK on their modulation. Comparing with our prototype, Sepetnitsky's prototype has more advantages, such as a nearer distance from the LED indicator to the camera and a higher frame rate of camera up to 60fps.

The contributions of our research are as follows:
\begin{enumerate}
\item It is difficult to build an available covert channel via the unmodulated LED status indicator. We proposed a novel approach on modulation form and presented a prototype to leak data from air-gapped network to an IP camera via the keyboard LED indicator.
\item A household IP camera with ordinary configuration is utilized to achieve the covert signal steadily in our experiment.
\end{enumerate}

The rest of the paper is organized as follows: Background technology is given in Section \ref{BackgroundTechnology}. A prototype, KLONC, is proposed in Section \ref{AttackModel}. Section \ref{ResultandEvaluation} presents results and evaluations. Countermeasures are given in Section \ref{Countermeasures}, and we draw our conclusions in Section \ref{Conclusions}.
\section{Background Technology}\label{BackgroundTechnology}
\subsection{LED}\label{LED}
A light-emitting diode (LED) is a two-lead semiconductor light source. It is a p-n junction diode that emits light when activated. When a suitable voltage is applied to the leads, electrons are able to recombine with electron holes within the device, releasing energy in the form of photons. 
\cite{wikiLED}

Most keyboards equip with three LED indicators recently. They are NumLock, CapsLock and ScrollLock arranged horizontally in the upper right corner of font panel.

\subsection{IP camera}
An IP camera\cite{wikiIPcamera}, also called a network surveillance camera, is a new type of camera which can access Internet. The user can control it by manipulating a client panel remotely. With development of the technology of optics , video coding and networks, the configuration of IP camera has upgraded rapidly. MPEG4 coding algorithm with h.264 standard is applied to cope with the high resolution up to 720P(1280x720) or 1080P(1920x1080). Nowadays, IP camera is widely used in normal life. 

\section{Attack Model}\label{AttackModel}
An attack model named KLONC, is proposed in this section. In the model, we suppose that the IP camera is compromised by an attacker. And the LED indicators on a keyboard of an air-gapped PC are in the camera's line of sight. We also suppose that a malware that controls the LEDs is preinstalled on the PC.

\begin{figure}
\includegraphics[width=0.5\textwidth]{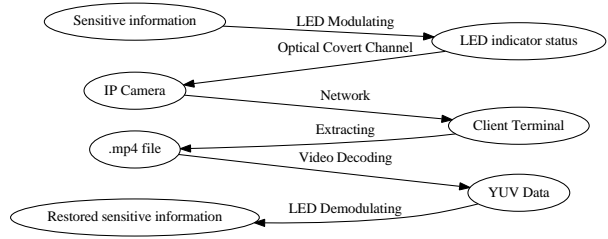} 
\caption{Flow diagram of KLONC}
\label{FlowdiagramofKLONC}
\end{figure}

As shown in Figure \ref{FlowdiagramofKLONC}, the sensitive information, such as credit card number, password, encryption key etc, exfiltrates via the LED indicator of the keyboard on the desk of an office cuibicle. The optical signal is fetched by an IP camera hung on the ceiling of the office. An optical covert channel is build between the LED indicator and the IP camera. Then, the attacker access the IP camera via Internet by manipulating the client panel of the IP camera with ID and password gotten forehand. A .mp4 video file is obtained by attacker. The YUV data of the LED indicator is gotten after decoding the video file. By demodulating the brightness values, the sensitive information is restored.

\subsection{Modulation and Encoding}
A normal method to leak messages is to turn the three LED indicators on/off on a keyboard. Then, the optical signals can be adopted by some type of acquisition equipment.

We can control those LEDs by \textit{keybd\underline{\hspace{0.5em}}event}() function\cite{APIkeybdEvent} in Windows API. The function synthesizes a keystroke. A hardware scan code is needed for the key. VK\underline{\hspace{0.5em}}NUMLOCK is the code for the key NumLock, and VK\underline{\hspace{0.5em}}CAPITAL or VK\underline{\hspace{0.5em}}SCROLL for the key CapsLock or ScrollLock. \textit{GetKeyState}() function\cite{APIGetKeyState} can be use to judge the LED's status. It returns 0 while LED turns off; It returns 1 while LED turns on. The function can be used to record LEDs' initial status to recovery their status after the covert signal transmitting.

The advantage of the method is threefold: A good compatibility for different Windows versions; Supporting both PS/2 and USB interfaces; No administrator privilege is required. The disadvantage is that the lock status of a LED indicator is changed while it is being turned on/off. Hence a interference would be made when the user is  typing in the meantime. Because this method on typing simulation is to send data into the keyboard buffer, reaction speed of LED indicators can be increased by modifying Registry keys for Windows.\cite{TechnetKeyboard}

For Linux OS, there are two methods to turn the LED indicators on/off. The command \textit{setleds} can turn them on/off without changing their lock statuses. But an administrator privilege is required. On the contrary, The command \textit{xset} and \textit{numlockx} can turn them on/off without any administrator privilege. But they change the lock statuses of the LEDs.

On modulation, the simplest form of a common modulation is On-Off Keying(OOK). We can use the presence of a signal(LED-ON) to encode a logical zero(0), and use the absence of a signal(LED-OFF) to encode a logical one(1).
\begin{center}
\begin{tabular}{c|c}
\hline
Logical Bit & LED Status\\
\hline
0 & LED-ON\\
1 & LED-OFF\\
\hline
\end{tabular}
\end{center}
The OOK can be used for the transmission with high carrier frequency. When the frequency is up to 150Hz\cite{Loughry:2002:ILO:545186.545189}, people can never find any flicker. But it is not available with low carrier frequency. More unfortunately, the frame rate of an IP camera is 15fps(Frames Per Second) at most. So we found a new form of signal modulation which is more suitable to transmit optical signal to a low-frequency acquisition equipment with a high covertness against human eyes.

\begin{figure}
\includegraphics[width=0.5\textwidth]{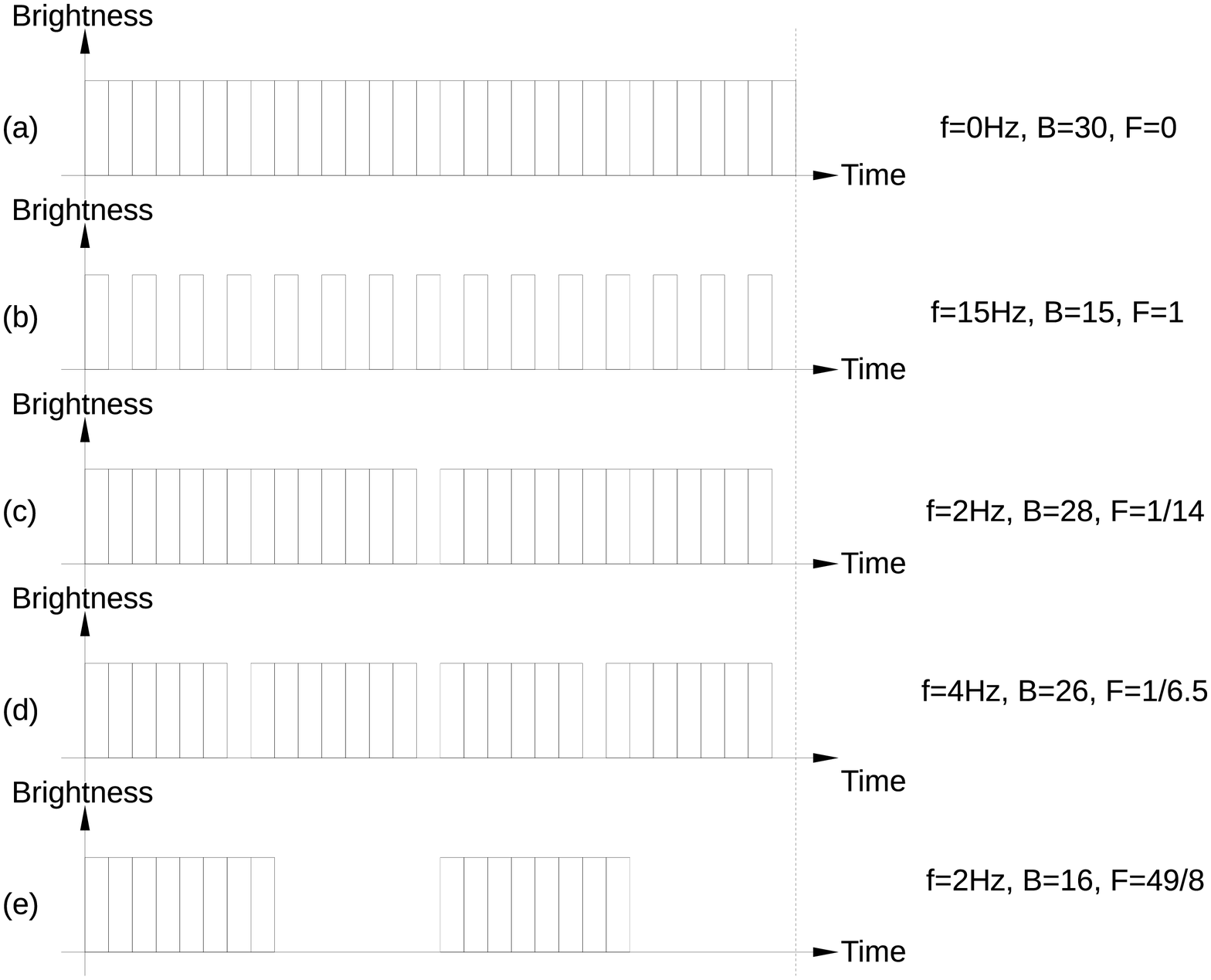} 
\caption{Frequency Simulations for B-FSK}
\label{FrequencySimulationsforBFSK}
\end{figure}

In our approach, we use Binary Frequency Shift Keying(B-FSK) to modulate the signal. We can use one flicker frequency $f_0$ to encode a logical zero(0), and use another flicker frequency $f_1$ to encode a logical one(1).
\begin{center}
\begin{tabular}{c|c}
\hline
Logical Bit & Flicker Frequency\\
\hline
0 & $f_0$\\
1 & $f_1$\\
\hline
\end{tabular}
\end{center}
Because there are only two discrete states on the brightness of a LED indicator, a novel method is proposed to simulate flicker frequencies on B-FSK. The method is described in Figure \ref{FrequencySimulationsforBFSK}.
 
In Condition (a) of Figure \ref{FrequencySimulationsforBFSK}, the LED is always being on, no flicker exists. Supposing the change rate is 30 times per second. So the flicker frequency $f=0$(No change happened), the brightness $B=30$(the sum of turn-on blocks), the flicker value $f$(the index to estimate human vision of flicker) is 0.

We define the \textbf{flicker value} $f$ by the following formula to express \textit{the feeling of flicker}.
\[f=\frac{D_{\text{off}}^2}{D_{\text{on}}}\]
Where, $D_{\text{off}}$ is the average length of the runs of turn-off block, $D_{\text{on}}$ is the average length of the runs of turn-on block. Obviously the Condition (e) is not good for covertness. It is the reason why OOK is not suitable to be a modulation form here. When a long runs of 1 follow a long runs of 0, the flicker value of the LEDs would be too high. The user would become aware of it.

The optical signal emitted from the LED is received by an IP camera. The video data are stored in a TF card inserted in the camera encoded by H.264 standard\cite{Team2013DraftH264}.
\subsection{Decoding and Demodulation}
We can use the famous free software \textit{FFmpeg} to convert the .mp4 video file into a .rgb video file with a command as follow.

\texttt{ffmpeg -i input.mp4 -vcodec rawvideo -pix\underline{\hspace{0.5em}}fmt rgb24 -an output.rgb}

But this method is not wise for the .rgb file will be too large. So we deal with the .mp4 file with following steps:
 
Firstly, the video data encoded by H.264 standard will be extracted from the .mp4 file.

\texttt{ffmpeg -i input.mp4 -f h264 output.264}

Secondly, the H.264 video will be decoded into YUV format data frame by frame. By referring to Lei's code\cite{Xiaohua}, we finished this step by making a C code with FFmpeg's \textit{avcodec} library.

Finally, pixel values of the LED indicator will be acquired from the YUV data by its fixed position(row and column) in the frame. There are three sample modes of YUV data: YUV444, YUV422 and YUV420. Take YUV420 for example, every pixel has a unique Y value, and four adjacent pixels share a set of U value and V value as shown in the Table \ref{YUV420sample}. So when the width of the frame is $w$, the pixel $(n,m)$'s offset in Y sequence is $(m-1)\times w + n$. And the offsets in U and V sequences are both $(\lfloor\frac{m+1}{2}\rfloor-1)\times\frac{w}{2}+\lfloor\frac{n+1}{2}\rfloor$.
\begin{table}
\centering
\caption{YUV420 sample}
\footnotesize
\begin{tabular}{|c|c|c|c|c|}
\hline 
$Y_{11}U_{11}V_{11}$ & $Y_{12}U_{11}V_{11}$ & $Y_{13}U_{12}V_{12}$ & $Y_{14}U_{12}V_{12}$ & $\cdots$ \\
\hline 
$Y_{21}U_{11}V_{11}$ & $Y_{22}U_{11}V_{11}$ & $Y_{23}U_{12}V_{12}$ & $Y_{24}U_{12}V_{12}$ & $\cdots$ \\
\hline 
$Y_{31}U_{21}V_{21}$ & $Y_{32}U_{21}V_{21}$ & $Y_{33}U_{22}V_{22}$ & $Y_{34}U_{22}V_{22}$ & $\cdots$ \\
\hline 
$Y_{41}U_{21}V_{21}$ & $Y_{42}U_{21}V_{21}$ & $Y_{43}U_{22}V_{22}$ & $Y_{44}U_{22}V_{22}$ & $\cdots$ \\
\hline 
$\cdots$ & $\cdots$ & $\cdots$ & $\cdots$ & $\cdots$ \\
\hline 
\end{tabular}
\label{YUV420sample}
\end{table}

As we mentioned on modulation, B-FSK is used as the modulation form. So naturally, we can demodulate the data by distinguishing two different frequencies.

In addition, we can calculate the mean value and the variance of the data. Because the every condition of Figure \ref{FrequencySimulationsforBFSK} has a B value, the index value of brightness, the mean value of Y value in data can be calculated to   distinguish two different B values. The variance of Y value in data can also represents the dither degree of the signal. It can be used to demodulate the data too.
\subsection{Effective Distance}
The effective distance is an essential index of a camera to fetch the optical signal of LED indicators. The ability of a camera is determined by its frame resolution and sensitivity of its electronics. So, on distances, there is an upper bound to obtain the message availably for a certain camera. Three factors influence the upper bound of effective distance. They are:
\begin{enumerate}
\item Ambient Brightness;
\item Emitting Angle of a LED indicator;
\item Distance between the LED indicator and the camera
\end{enumerate}
\subsubsection{Ambient Brightness}
LED indicators are only used to represent the statuses of a keyboard, so the brightness of a LED indicator is always weak. When ambient brightness is too high, the brightness status of a LED can hardly be distinguished in video. On the contrary, When ambient brightness is low enough, the status of a LED is quite obvious in video.

Nevertheless, in our experiments, we find that when the camera is close to the keyboard, a certain intensity of ambient brightness can reduce the noise in MPEG-4 video, the capacity of the channel is increased instead.
\subsubsection{Relationship between Emitting Angle and Distance}
The \textbf{emitting angle of a LED indicator} is defined here as \textit{an angel of the direction of LED's emitting and the direction of the camera}.

\begin{figure}
\includegraphics[width=0.5\textwidth]{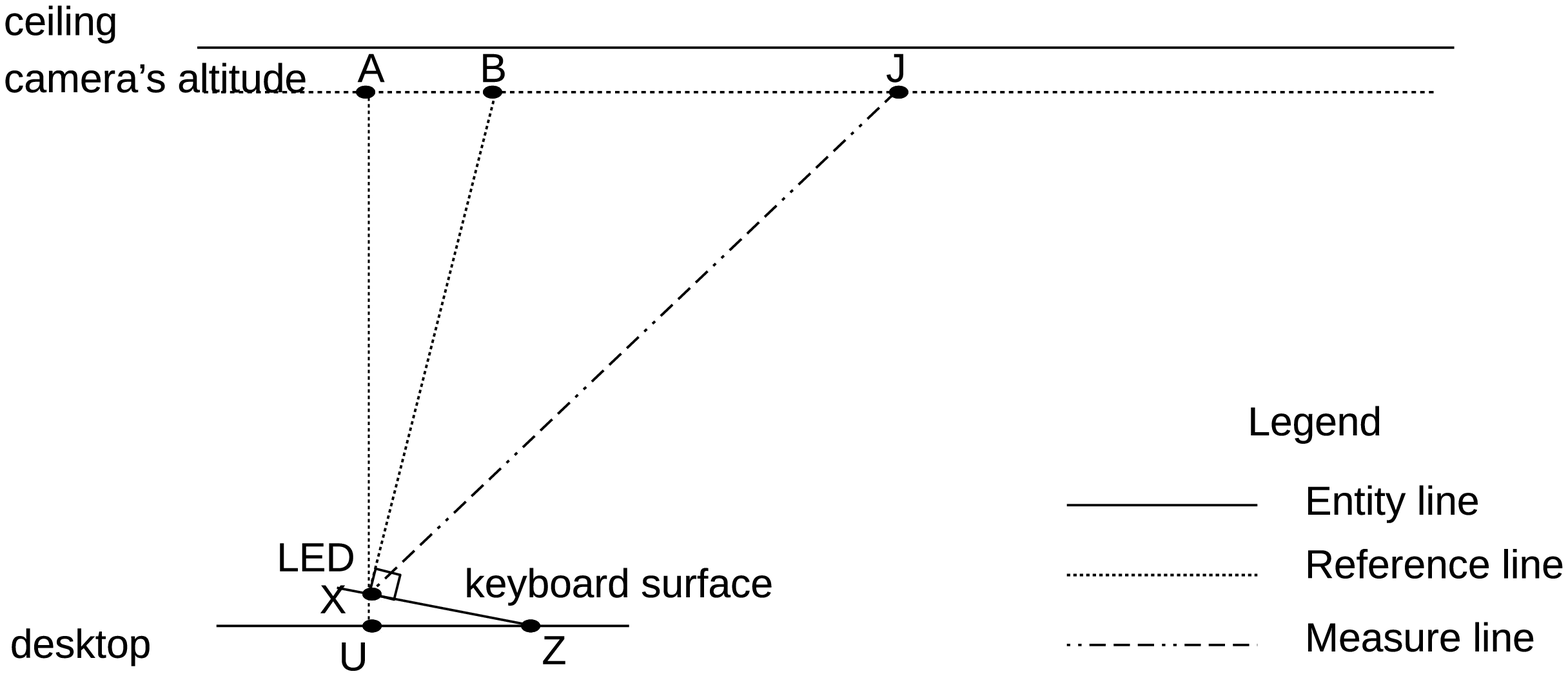} 
\caption{Relationship between Emitting Angle and Distance}
\label{RelationshipbetweenEmittingAngleandDistance}
\end{figure}

IP cameras are always hung on the ceiling. The distance between the desktop and the ceiling is constant. So, the longer the distance from LED indicator to camera, the bigger the emitting angle, the weaker the intensity of signal obtained. The relationship between the emitting angle and the distance is described in Figure \ref{RelationshipbetweenEmittingAngleandDistance}. Where, $\angle \text{UZX}$ is the angle between keyboard surface and desktop. According to the feet's status of keyboard, the $\angle \text{UZX}$ has two fixed value. Taking Logitech K120 as an example, the values of $\angle \text{UZX}$ are $1.1353328^{\circ}$ and $6.9474259^{\circ}$. 

In general, LED's emitting direction is perpendicular to the surface of keyboard. That is $\text{XB}\perp\text{XZ}$, then $\angle \text{UZX} = \angle \text{AUB}$.
Suppose J is an arbitrary plot on the line AF, then we can get a relational expression between the emitting angle and the distance as follow.
\[\angle\text{JXB}=\arccos\left(\frac{|\text{XA}|}{|\text{XJ}|}\right)-\angle\text{UZX}\]
Where, $\angle \text{UZX} = \angle \text{AUB}$ is known, and $|\text{XA}|$ can be measured. 

We can also determine the value of LED's visual angle in camera, and the value of LED's effective shine area in the projection plane.

\subsubsection{Relationship between Distance and Brightness}
\begin{figure}
\includegraphics[width=0.5\textwidth]{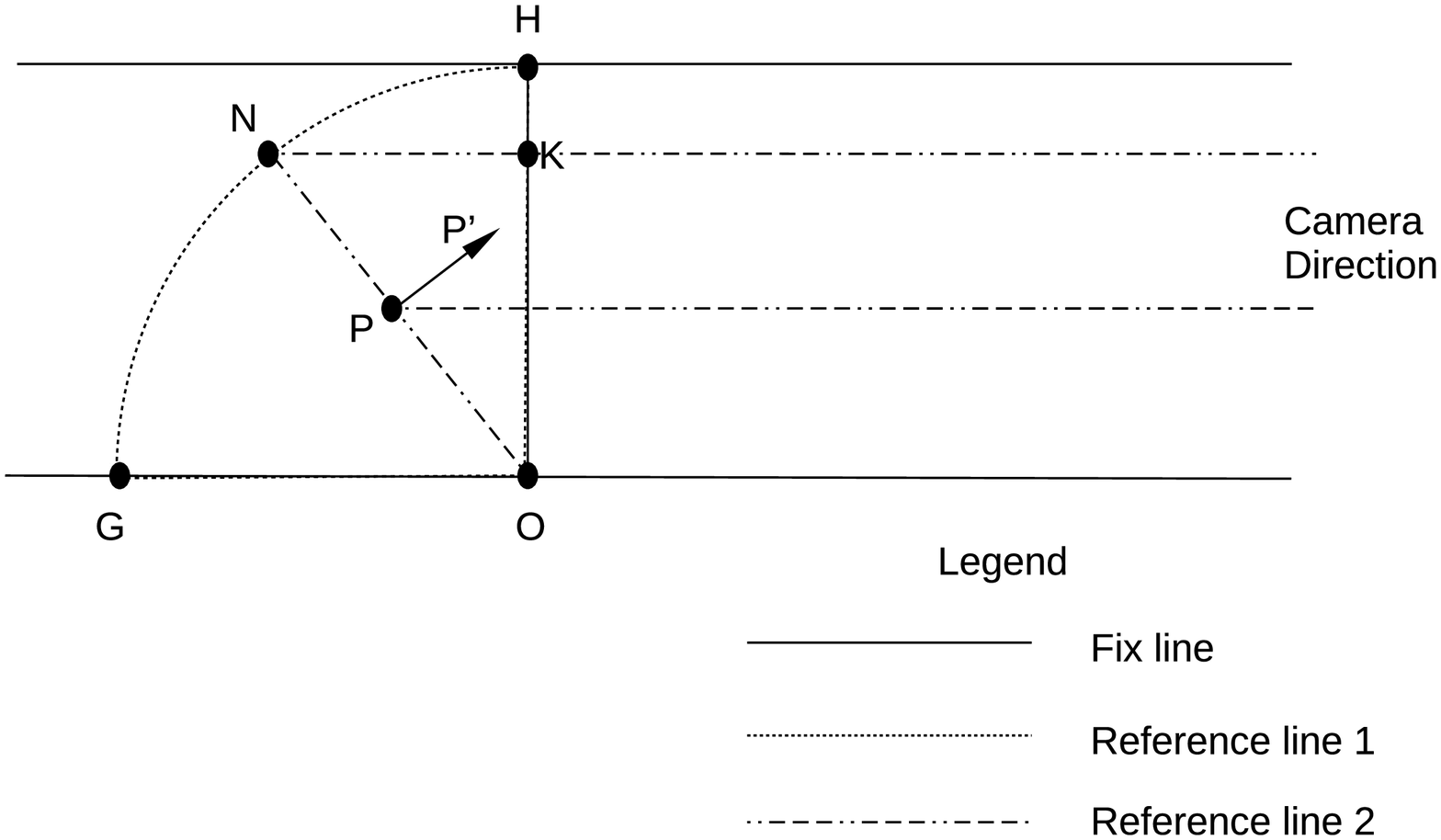} 
\caption{Relationship between Emitting Angle and Effective Shine Area}
\label{RelationshipbetweenEmittingAngleandEffectiveShineArea}
\end{figure}

Because $\frac{|\text{OY}|}{|\text{ON}|}\approx 1000$, it means that$\angle\text{HOY}=\arccos\left(\frac{|\text{ON}|}{2|\text{OY}|}\right)$ is approximated with $90^\circ$. So, a simplified model is described in Figure \ref{RelationshipbetweenEmittingAngleandEffectiveShineArea}.
In the figure, Plot O is one side of the LED indicator. Plot N is the other side. Changing with the emitting angle from $0^\circ$ to $90^\circ$, N moves on the Arc GH. Plot K is N's projection on the camera direction. Then $|\text{KO}|$ can be represent the value of LED's effective shine area
in the projection plane. We can get a relational expression between the emitting angle and the effective shine area as follow.
\[|\text{KO}| = |\text{ON}|\cos(\angle\text{HON})\]
Where, $\angle\text{HON}$ is equal to the emitting angle.

Furthermore, the brightness in the video is not only related with the effective shine area, but also with the distance between LED indicator and camera. Then, the relational expression between the brightness and its influence factors is:
\[B=\beta\frac{|\text{ON}|\cos(\angle\text{HON})}{\left(\frac{|\text{OY}|}{y}\right)^2}\]
Where, $\beta$ is a constant coefficient, $y$ is an initial reference distance, a non-zero value. $|\text{ON}|$ is the length of LED indicator in the direction of change.

\subsection{Channel Capacity}
According to Nyquist-Shannon sampling theorem, if the sampling frequency of the receiver is $f$, the maximum carrier frequency would be $\frac{f}{2}$. The frame rate of most normal camera in current market is 25fps(frames per second). It means an upper bound of transmitting speed. The frame rate of some high end camera can be 60fps or higher. But high frame rate and high resolution are interacted on each other. For example, the frame rate of most IP camera is 25fps in 720P, but 15fps in 1080P. Aiming at surveillance for security, there is no tend to increase the frame rate of IP camera.
\subsection{Covertness}
According to the persistence of vision\cite{wikiPersistenceofvision}, a single slight change in 50ms(microsecond) is not sensitive to human vision. This feature can help us to hide a turn-off behavior in 40ms on a LED indicator always being on. For mankind, the maximal fusion frequency can be up to 60Hz at very high illumination intensities \cite{wikiFlickerfusionthreshold}.By conducting experiments, we find that a turn-off behavior in 20ms can hardly is observed even the LED is stared continuously. When the duration of the turn-off behavior is in 20ms to 50ms, a tiny dithering on the brightness of LED can be observed under a careful observation.

Moreover, the covertness of three LED indicators on the keyboard are different. To normal computer users, they would suspect something wrong with their computers when they notice that a LED indicator turns on without any sake, even when they find any tiny flash on the brightness of a LED. So our only choice is to select the LED indicator always being on to leak data covertly.

Among three LED indicators, NumLock is always on after a booting of Windows on most computers. Hence NumLock is most suitable to leak covert message, unless on the computers in department of finance where the number pads will be served all the time. ScrollLock is another suitable one actually for its function is too old to current OSes. If ScrollLock could keep being on from the booting of Windows, it would not catch the attention of the user. On the contrary, the function of CapsLock is alway used by every user to input text message such as ID, password etc. So it would make user anxious when the turn-on CapsLock is seen.

\section{Results and Evaluations}\label{ResultandEvaluation}
\subsection{Experiment Setting}
\begin{figure}
\includegraphics[width=0.5\textwidth]{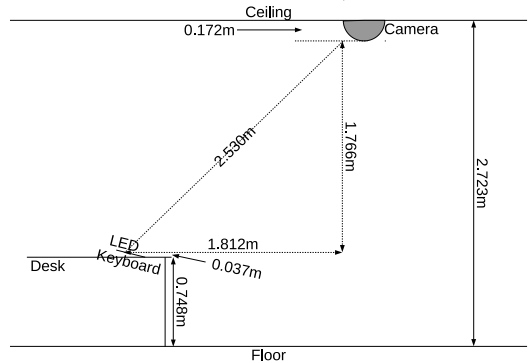} 
\caption{Survey Sheet of Experimental Environment}
\label{ExperimentalEnvironment}
\end{figure}

An open-plan office is served as the experimental environment. It is a common environment for most business companies and research organizations etc. The keyboard that leaks data is located on the desk of an office cubicle. The IP camera is hung on the ceiling of the office. A survey sheet of the experimental environment is shown in Figure \ref{ExperimentalEnvironment}.

The configuration lists of Personal Computer and IP camera are shown in Table \ref{ConfigurationofPersonalComputer} and Table \ref{ConfigurationofIPCamera}.
\begin{table*}
\caption{Configuration of Personal Computer}
\small\centering
\begin{tabular}{c|c}
\hline
Module & Configuration \\
\hline
CPU & Intel Core i5-4590 CPU 3.30GHz\\
Motherboard & ASUS B85-PLUS R2.0\\
RAM & 8GB\\
Hard Disk & SEAGATE Desktop HDD 500G\\
Keyboard & Logitech K120 HID USB\\
OS &Windows 10 Chinese Simplified Version 64-bit (10.0, Build 14393)\\
\hline
\end{tabular}
\label{ConfigurationofPersonalComputer}
\end{table*}

\begin{table*}
\caption{Configuration of IP Camera}
\small\centering
\begin{tabular}{c|c}
\hline
Module & Configuration\\
\hline
Resolution & 1920x1080 and 640x352\\
Video Encoding & H264MANINPROFLE, JPEG Snapshot\\
Wireless Network & IEEE 802.11b/g/n 2.4GHz\\
Focus & 5 times optical zoom, 3.6-12mm\\
Aperture value & F2.0\\
\hline
\end{tabular}
\label{ConfigurationofIPCamera}
\end{table*}

\subsection{Results}
Several experiments were conducted with different distances and various ambient brightness. Obtained BERs(Bit Error Rates) of the covert channel KLONC are listed in Table \ref{BER}. The table shows that BERs increase with the distances, but are not linear relationships with the ambient brightness. When the distance is 2.54m, most of BERs are less than 10\%. When the distance is 3.27m, most of BERs are less than 25\%. But we can see all BERs are greater than 33\% while the distance reaches 5 meters. 

According to the channel capacity formula in Information Theory:
\[C=1-H(p)=1+p\log(p)+(1-p)\log(1-p)\]
we know that when $p=\frac{1}{3}$, the capacity $C=0.081704166<\frac{1}{12}$. It means that we need more than 12 bits data to transmit 1 bit information correctly. It is impossible to build a reliable channel under such a condition. So, the distance 5 meters can be considered as an upper bound of effective distance to build a covert channel with current experimental devices.
\begin{table*}
\caption{Bit Error Rates(\%) with Different Distances and Various Ambient Brightness}
\small
\begin{tabular}{c|cccccccccccc}
\hline 
Brightness(LUX) & 100 & 200 & 300 & 400 & 500 & 600 & 700 & 800 & 900 & 1000 & 1100 & 1200\\
\hline 
2.54m & 0.39 & 15.63 & 0 & 0 & 0 & 10.16 & 4.30 & 0 & 16.02 & 5.47 & 1.17 & 8.98\\
3.27m & 3.13 & 1.95 & 27.73 & 23.05 & 14.06 & 6.64 & 10.94 & 16.80 & 10.55 & 42.19 & 10.55 & 23.38\\
4.02m & 26.17 & 35.55 & 30.08 & 26.17 & 37.89 & 33.20 & 28.13 & 24.61 & 30.08 & 30.86 & 39.84 & 39.84\\
5.08m & 38.67 & 37.89 & 33.98 & 37.11 & 33.20 & 41.41 & 41.41 & 44.92 & 38.28 & 41.41 & 42.58 & 39.06\\
\hline 
\end{tabular}
\label{BER}
\end{table*}

In our experiments,  $\angle \text{UZX} = 6.9474259^\circ$(the angle between keyboard surface and desktop) and  $|\text{XA}|=1.77m$(the distance between LED and camera's altitude) in Figure \ref{RelationshipbetweenEmittingAngleandDistance}. A list of emitting angles and distances are given in Table \ref{EmittingAnglesandDistancesinExperiments}. We can see that the emitting angle grows up observably in the distances from 2.54m to 5.08m. It means that the camera receives a quick drop in brightness when the emitting angle increases.
\begin{table}
\caption{Emitting Angles and Distances}
\centering
\small
\begin{tabular}{c|cccc}
\hline
 & Exp.1 & Exp.2 & Exp.3 & Exp.4\\
\hline
Distance & 2.54m & 3.27m & 4.02m & 5.08m\\
Angle & \scriptsize$38.877^\circ$ & \scriptsize$50.2814^\circ$ & \scriptsize$56.9296^\circ$ & \scriptsize$62.6616^\circ$\\
\hline
\end{tabular}
\label{EmittingAnglesandDistancesinExperiments}
\end{table}

\begin{figure}
\includegraphics[width=0.5\textwidth]{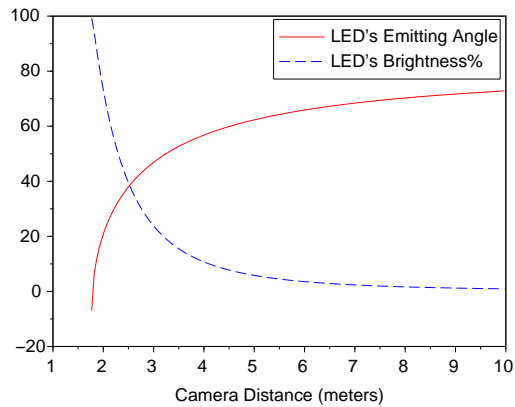} 
\caption{Relationship between Emitting Angle and Brightness}
\label{RelationshipbetweenEmittingAngleandBrightness}
\end{figure}

Then, a relational graph between distance, emitting angle and brightness  is made with  $y=1.77m$(the distance between LED and camera's altitude),  $\beta=1$(the constant coefficient) and  $|\text{ON}|=1$(the length of LED indicator in the direction of change) in Figure \ref{RelationshipbetweenEmittingAngleandBrightness}. The figure shows that the brightness captured by camera at a distance of 4 meters is about only 10\% of the brightness at 1.77 meters.

\subsection{Comparison with OOK}
When the flicker frequency is so low that the turn-off behavior can be found with human vision, it is natural to do something making the behavior more indetectable. A normal way is to give a long turn-on duration before a turn-off behavior. Then we can encode the message before modulation like this:
\begin{center}
\begin{tabular}{c|c}
\hline
Plain Bit & Encoded Word\\
\hline
0 & 0\\
1 & $0\cdots 01$\\
\hline
\end{tabular}
\end{center}
Then the channel rate $R_\text{OOK}$ and the flicker value $f_\text{OOK}$ can be deduced as follows.
\[R_\text{OOK}=\frac{2F}{|\text{Enc}(1)|+1},\quad f_\text{OOK}=\frac{1}{|\text{Enc}(1)|}\]
Where, $F$ is the flicker frequency, $\text{Enc}(1)$ is the length of encoded word of the bit one.

Meanwhile, the channel rate $R_\text{B-FSK}$ and the flicker value $f_\text{B-FSK}$ with $f_0=0$ can be deduced as follows.
\[R_\text{B-FSK}=f_1,\quad f_\text{B-FSK}=\frac{1}{\frac{2F}{f_1}-0.5}\]

\begin{figure}
\includegraphics[width=0.5\textwidth]{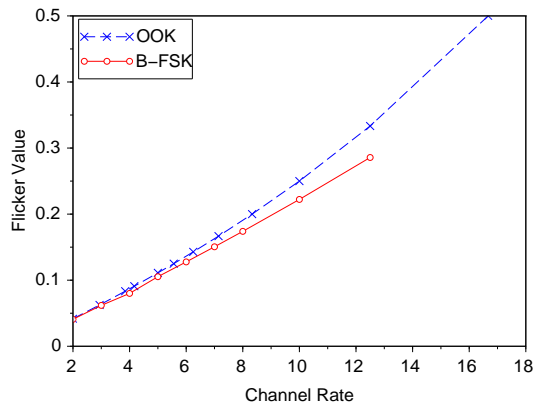} 
\caption{Flicker Values Comparison between B-FSK and OOK}
\label{B-FSKvsOOK}
\end{figure}

A comparison of flicker values is given in Figure \ref{B-FSKvsOOK}  with $F=25$ as same as Sepetnitsky's propotype\cite{Sepetnitsky-2014-6975588}. The figure shows that the flicker value with B-FSK is always lower than those with OOK.

\section{Countermeasures}\label{Countermeasures}
The countermeasures can be divided into two types: procedural countermeasures and technical countermeasures.

Procedural countermeasures involve banning cameras from the office, covering the LEDs, cutting off the LEDs' feet and shielding windows.
Any banning policy needs a supervision all the time to insure no exception. Covering the LEDs or cutting off the feet of LEDs is easy to utilized, but it makes users inconvenient without any indication. In addition, armored glass is used as walls in many office space. So a surveillance camera can also received optical signal through glass of the windows or wall. It is necessary to shield them availably.

Technical countermeasures involve LED status monitoring with software or optical methods, LED status confusing with software. Detecting the malware is a common job to security software. Then a watchdog for the status of LEDs can find the abuses on them. As a cost, CPU resources would be occupied to slow down the OS. Detecting the abuse of LEDs by an outside sensor is a perfect method without giving any information to the attacker. It always obtains a high percentage of success if the hardware meets the conditions. But the existence of a covert channel is a low probability event. So it is still difficult to detect. We notice that there is only one covert channel that can be established in same time. So we can confuse the LED status actively to hold back the real risk.

The list of all countermeasures is summarized in Table \ref{CountermeasuresList}.

\begin{table*}
\caption{Cost and Effect of Countermeasures}
\label{CountermeasuresList}
\begin{tabular}{l|c|c|c|l}
\hline 
Countermeasure & Type & Cost & Effect & shortcomings\\
\hline 
Banning cameras from the office & Proc. & High & Good & Need for supervision \\
Covering the LEDs & Proc. & Low & Good & Inconvenience to user\\
Cutting off the LEDs' feet & Proc. & Low & Good & Inconvenience to user\\
Shielding windows & Proc. & High & Good & Change surrounding brightness\\
\hline
Status monitoring with software & Tech. & Low & Good & Occupy CPU resources\\
Status monitoring with optical methods & Tech. & High & Normal & Difficult to detect\\
Status confusing with software & Tech. & Low & Good & Occupy CPU resources\\
\hline 
\end{tabular}
\end{table*}
\section{Conclusions}\label{Conclusions}
A novel form of signal modulation with the fix status LED indicator to build an optical covert channel was proposed in this paper. By using this modulation form, a LED indicator in Type I can leak covert signal with a good covertness on human vision. An attack model, KLONC, was given to build a covert communication with a purchasable  generally configured IP camera by programming C codes to turn the LED on/off. Furthermore, the modulation form and the corresponding demodulation method were designed and optimized. Then the efficiency and covertness were estimated. The upper bound of effective distance of KLONC was obtained with both theoretical calculation and experimental observation. Finally, countermeasures were given by considering the necessary conditions of existence of this kind of covert channel. 

\bibliographystyle{plain}

\begin{thebibliography}{10}

\bibitem{bartolini2016capacity}
Davide~B Bartolini, Philipp Miedl, and Lothar Thiele.
\newblock On the capacity of thermal covert channels in multicores.
\newblock In {\em Proceedings of the Eleventh European Conference on Computer
  Systems}, page~24. ACM, 2016.

\bibitem{guri2015gsmem}
Mordechai Guri, Assaf Kachlon, Ofer Hasson, Gabi Kedma, Yisroel Mirsky, and
  Yuval Elovici.
\newblock Gsmem: data exfiltration from air-gapped computers over gsm
  frequencies.
\newblock In {\em 24th USENIX Security Symposium (USENIX Security 15)}, pages
  849--864, 2015.

\bibitem{guri2014airhopper}
Mordechai Guri, Gabi Kedma, Assaf Kachlon, and Yuval Elovici.
\newblock Airhopper: Bridging the air-gap between isolated networks and mobile
  phones using radio frequencies.
\newblock In {\em 2014 9th International Conference on Malicious and Unwanted
  Software: The Americas (MALWARE)}, pages 58--67. IEEE, 2014.

\bibitem{guri2016usbee}
Mordechai Guri, Matan Monitz, and Yuval Elovici.
\newblock Usbee: Air-gap covert-channel via electromagnetic emission from usb.
\newblock In {\em 2016 14th Annual Conference on Privacy, Security and Trust
  (PST)}, pages 264--268. IEEE, 2016.

\bibitem{guri2015bitwhisper}
Mordechai Guri, Matan Monitz, Yisroel Mirski, and Yuval Elovici.
\newblock Bitwhisper: Covert signaling channel between air-gapped computers
  using thermal manipulations.
\newblock In {\em IEEE 28th Computer Security Foundations Symposium (CSF),
  2015}, pages 276--289. IEEE, 2015.

\bibitem{Guri-Fansmitter-2016arXiv160605915G}
Mordechai Guri, Yosef Solewicz, Andrey Daidakulov, and Yuval Elovici.
\newblock Fansmitter: Acoustic data exfiltration from (speakerless) air-gapped
  computers.
\newblock {\em arXiv preprint arXiv:1606.05915}, 2016.

\bibitem{Guri2017DiskFiltration}
Mordechai Guri, Yosef Solewicz, Andrey Daidakulov, and Yuval Elovici.
\newblock {\em Acoustic Data Exfiltration from Speakerless Air-Gapped Computers
  via Covert Hard-Drive Noise (`DiskFiltration')}, pages 98--115.
\newblock Springer International Publishing, Cham, 2017.

\bibitem{guri2017xled}
Mordechai Guri, Boris Zadov, Andrey Daidakulov, and Yuval Elovici.
\newblock xled: Covert data exfiltration from air-gapped networks via router
  leds.
\newblock {\em arXiv preprint arXiv:1706.01140}, 2017.

\bibitem{Guri2017LED}
Mordechai Guri, Boris Zadov, and Yuval Elovici.
\newblock {\em LED-it-GO: Leaking (A Lot of) Data from Air-Gapped Computers via
  the (Small) Hard Drive LED}, pages 161--184.
\newblock Springer International Publishing, Cham, 2017.

\bibitem{hanspach2014covert}
Michael Hanspach and Michael Goetz.
\newblock On covert acoustical mesh networks in air.
\newblock {\em Journal of Communications}, 8(11):758--767, 2013.

\bibitem{kuhn1998soft}
Markus~G Kuhn and Ross~J Anderson.
\newblock Soft tempest: Hidden data transmission using electromagnetic
  emanations.
\newblock In {\em International Workshop on Information Hiding}, pages
  124--142. Springer, 1998.

\bibitem{Lampson:1973:NCP:362375.362389}
Butler~W. Lampson.
\newblock A note on the confinement problem.
\newblock {\em Commun. ACM}, 16(10):613--615, October 1973.

\bibitem{lee2015various}
Eunchong Lee, Hyunsoo Kim, and Ji~Won Yoon.
\newblock Various threat models to circumvent air-gapped systems for preventing
  network attack.
\newblock In {\em International Workshop on Information Security
  Applications(WISA)}, pages 187--199. Springer, 2015.

\bibitem{Xiaohua}
Xiaohua Lei.
\newblock Simplest ffmpeg decoder pure.
\newblock \url{http://blog.csdn.net/leixiaohua1020/article/details/42181571},
  2015.
\newblock [Online; accessed 30-September-2017] Chinese Language.

\bibitem{lopes2017platform}
Arthur~Costa Lopes and Diego~F Aranha.
\newblock Platform-agnostic low-intrusion optical data exfiltration.
\newblock In {\em International Conference on Information Systems Security \&
  Privacy(ICISSP)}, pages 474--480, 2017.

\bibitem{Loughry:2002:ILO:545186.545189}
Joe Loughry and David~A. Umphress.
\newblock Information leakage from optical emanations.
\newblock {\em ACM Trans. Inf. Syst. Secur.}, 5(3):262--289, August 2002.

\bibitem{masti2015thermal}
Ramya~Jayaram Masti, Devendra Rai, Aanjhan Ranganathan, Christian M{\"u}ller,
  Lothar Thiele, and Srdjan Capkun.
\newblock Thermal covert channels on multi-core platforms.
\newblock In {\em 24th USENIX Security Symposium (USENIX Security 15)}, pages
  865--880, 2015.

\bibitem{Mirsky2017}
Y.~Mirsky, M.~Guri, and Y.~Elovici.
\newblock Hvacker: Bridging the air-gap by manipulating the environment
  temperature.
\newblock {\em Magdeburger Journal zur Sicherheitsforschung}, 14:815--829,
  August 2017.
\newblock Retrieved August 18, 2017.

\bibitem{APIGetKeyState}
MSDN.
\newblock Getkeystate function (windows).
\newblock
  \url{https://msdn.microsoft.com/en-us/library/windows/desktop/ms646301(v=vs.85).aspx}.
\newblock [Online; accessed 18-September-2017].

\bibitem{APIkeybdEvent}
MSDN.
\newblock keybd\underline{\hspace{0.5em}}event function (windows).
\newblock \url{https://msdn.microsoft.com/en-us/library/ms646304(VS.85).aspx}.
\newblock [Online; accessed 18-September-2017].

\bibitem{Malley-o2014bridging}
Samuel~Joseph O’Malley and Kim-Kwang~Raymond Choo.
\newblock Bridging the air gap: Inaudible data exfiltration by insiders.
\newblock 2014.

\bibitem{selber2017uncovert3}
Mirko Selber and Prof Dr~Lothar Thiele.
\newblock Uncovert3: Covert channel attacks on commercial multicore systems.
\newblock 2017.

\bibitem{Sepetnitsky-2014-6975588}
V.~Sepetnitsky, M.~Guri, and Y.~Elovici.
\newblock Exfiltration of information from air-gapped machines using monitor's
  led indicator.
\newblock In {\em 2014 IEEE Joint Intelligence and Security Informatics
  Conference}, pages 264--267, Sept 2014.

\bibitem{shamir2014light}
Adi Shamir.
\newblock Light-based printer attack overcomes air-gapped computer security.
\newblock
  \url{https://www.scmagazineuk.com/light-based-printer-attack-overcomes-air-gapped-computer-security/article/541140/},
  2014.
\newblock UK, SG SC Magazine,[Online; accessed 18-September-2017].

\bibitem{Team2013DraftH264}
Joint~Video Team.
\newblock Draft itu-t recommendation and final draft international standard of
  joint video specification.
\newblock 2013.

\bibitem{TechnetKeyboard}
Microsoft Technet.
\newblock keyboard.
\newblock \url{https://technet.microsoft.com/en-us/library/cc978656.aspx},
  2017.
\newblock [Online; accessed 30-September-2017].

\bibitem{wikiFlickerfusionthreshold}
Wikipedia.
\newblock Flicker fusion threshold --- {W}ikipedia{,} the free encyclopedia.
\newblock \url{https://en.wikipedia.org/wiki/Flicker_fusion_threshold}, 2017.
\newblock [Online; accessed 30-September-2017].

\bibitem{wikiIPcamera}
Wikipedia.
\newblock Ip camera --- {W}ikipedia{,} the free encyclopedia.
\newblock \url{https://en.wikipedia.org/wiki/IP_camera}, 2017.
\newblock [Online; accessed 18-September-2017].

\bibitem{wikiLED}
Wikipedia.
\newblock Light-emitting diode --- {W}ikipedia{,} the free encyclopedia.
\newblock \url{https://en.wikipedia.org/wiki/Light-emitting_diode}, 2017.
\newblock [Online; accessed 18-September-2017].

\bibitem{wikiPersistenceofvision}
Wikipedia.
\newblock Persistence of vision --- {W}ikipedia{,} the free encyclopedia.
\newblock \url{https://en.wikipedia.org/wiki/Persistence_of_vision}, 2017.
\newblock [Online; accessed 30-September-2017].

\bibitem{Zander2007-4317620}
S.~Zander, G.~Armitage, and P.~Branch.
\newblock A survey of covert channels and countermeasures in computer network
  protocols.
\newblock {\em IEEE Communications Surveys Tutorials}, 9(3):44--57, Third 2007.

\end{thebibliography}

\end{document}